 \documentclass[preprint2]{aastex}

\begin{document}


\title{Cosmic Star Formation History to z=1 from a Narrow Emission Line Selected Tunable Filter Survey }


\author{Karl Glazebrook{\altaffilmark{1}}, Jeffrey Tober{\altaffilmark{1}},  Scott Thomson{\altaffilmark{2}}, 
Joss Bland-Hawthorn{\altaffilmark{2}}, Roberto Abraham{\altaffilmark{3}}}


\altaffiltext{1}{Department of Physics and Astronomy, Johns Hopkins University, Baltimore, MD 21218}
\altaffiltext{2}{Anglo-Australian Observatory, P.O.Box 296, Epping NSW, Australia}
\altaffiltext{3}{Department of Astronomy, University of Toronto, Toronto,ON M5S3H8, Canada}


\begin{abstract}
We report the results of a deep 3D imaging survey of the Hubble Deep Field North using the Taurus Tunable Filter and the William Herschel Telescope. This survey was designed to search for new line emitting populations of objects missed by other techniques and to measure the cosmic star-formation rate density from a line-selected survey. We observed in three contiguous sequences of narrow band slices in the 7100, 8100 and 9100\AA \space regions of the spectrum, corresponding to a cosmological volume of up
to 1000 Mpc$^3$ at $z=1$, down to a flux limit of $\sim 2\times 10^{-17}$ ergs cm$^{-2}$ s$^{-1}$. The survey is deep enough to be highly complete for low line luminosity galaxies.
Cross-matching with existing spectroscopy in the field results in a small line-luminosity limited sample, with very highly redshift identification completeness containing seven [OII], H$\beta$ and H$\alpha$ emitters over the redshift range 0.3 -- 0.9. Treating this as a direct star-formation rate selected sample we estimate the star-formation history of the Universe to $z=1$. We find  no evidence for any new population of line emitting objects contributing significantly to the cosmological star-formation rate density. Rather from our complete narrow-band sample we find the star-formation history is consistent with earlier estimates from broad-band imaging surveys and other less deep line-selected surveys.
\end{abstract}



\section{Introduction}

\def\todo#1{{[\bf TODO: #1]}}

Studies of cosmic star formation history over the past decade have been an integral part of the development of our knowledge of galaxy formation and evolution.  These studies have found that star formation rates rise significantly from z = 0 to z = 1, peak between 1 $<$ z $<$ 2, and flatten or decline (depending on the role of
dust obscuration on ultraviolet measures) beyond z = 2 \citep{mad97,conn97,stei99,kgb99,flor99,gia04}.  Star formation rates (SFR) are usually determined by measuring the luminosity densities of galaxies, either from emission lines or from the continuum spectrum of a wavelength range.  Highly luminous stars live very short lifetimes, and can be used as a proxy for instantaneous star formation rates in galaxies.  These stars easily ionize the gases surrounding them, and the strength of the recombination lines of various elements gives the number of ionizing photons, which can be converted into star formation rates \citep{ken92}. They also 
give rise to broad-band ultraviolet emission which can also be used as 
a measure of star-formation rate accessible at high-redshift \citep{mad97}, but this should be used with
caution as the ultraviolet is much more 
subject to dust extinction than the optical re-combination lines and high-redshift SFRs can be seriously underestimated \citep{kgb99,stei99}.

In the intermediate redshift $0<z<1$ regime
the most commonly studied emission lines for measuring the cosmological SFR 
are H$\alpha$ \citep{gal95, tre98, kgb99, tre02, kew02, fuj03}, and the 3727\AA \space OII line \citep{hog98, kew04}.  H$\alpha$, when corrected for extinction is directly proportional to
the number of ionizing photons from massive stars \citep{oster89}. Since such stars are short-lived ($\sim$ 20 Myr) their abundance traces the instantaneous SFR. H$\beta$ can also be used this way
but the extinction and stellar absorption corrections are larger. The [OII] emission line (3727\AA) has a more complex dependence
on metallicity  \citep{kew04} and is less straight-forward to use as a SFR tracer.

Most studies use broad band filters to select a magnitude and/or color-limited samples which gives rise to selection effect issues when computing the total SFR. For example populations of strong-lined high equivalent width objects may contribute significantly to the SFR while having weak continuum emission and hence lying below typical broad-band magnitude limits. An alternative approach is to use a narrow-band filter to directly select emission line objects down to an emission line flux limit. So far as the line luminosity is in proportion to the SFR then this is equivalent to a SFR-selected sample and if the line's redshifted wavelength matches that of the narrow filter's one obtains a higher signal:noise on lower luminosity objects in a given integration time. Also, this technique produces a volume-limited sample, since the narrow observed bands correspond to small windows in redshift space with approximately constant luminosity limits.

Previous narrow-band filter surveys have almost universally relied on using special interference filters. Some of the deepest,
most recent work includes:   \citet{pas01} who surveyed 684  arcmin$^{2}$ 
to a flux limit of about 5 $\times$ 10$^{-16}$ erg s$^{-1}$ cm$^{-2}$, 
finding 16 objects at $z = 0.24$ and  \citet{moo02} surveyed 100 arcmin$^{2}$ in the near-infrared to a flux limit 
of  5-12 $\times$ 10$^{-17}$ erg s$^{-1}$ cm$^{-2}$, finding 6 objects at $z = 2.2$. A recent target of these
searches has been high-$z$ Lyman-$\alpha$ emission: \cite{stei00} surveyed 94 arcmin$^2$ in 
a deep Keck exposure reaching fluxes of $4\times$10$^{-17}$ erg s$^{-1}$ cm$^{-2}$; \cite{Hu04} extending
earlier work \citep{Hu98,Hu99}  surveyed 0.25  deg$^2$ to a depth of $2\times$10$^{-17}$ erg s$^{-1}$ cm$^{-2}$ 
and  \cite{LALA} reached the same flux limit  over 0.72 deg$^2$; however spectroscopic follow-up of these surveys
has been limited to $z>3$ objects (winnowed by either broad-band color and/or high-equivalent width selection).

A novel approach to selecting line emitters at high-redshift is to use a {\it tunable\/} filter, of which the most practical 
technology to realise this is based on the Fabry-Perot etalon \citep{FPref}. Tunable filters have unique advantages starting
with the fact 
that a contiguous scan can be made in wavelength/redshift space allowing a line to be unambiguously identified via a peak in 
flux without reference to any broad-band image. They also offer much higher resolution ($\Delta\lambda\sim 10\,$\AA) than
interference filters  ($\Delta\lambda\sim 100\,$\AA) which better match the typical line widths of galaxies with consequent
signal:noise gains. The scan through the line also permits any continuum to be precisely removed in the analysis allowing accurate integration through the line profile to measure the flux. This flux can also be integrated over the entire  galaxy (essentially 3D aperture photometry) meaning total line fluxes can easily be measured without the dubious aperture corrections required
for narrow slit spectra.

 There have been few surveys using this technique. \citet{jon01} using the Taurus Tunable Filter 
on the Anglo-Australian Telescope 
surveyed 972 arcmin$^{2}$ to a flux limit of 5--10 $\times$ 10$^{-17}$ erg s$^{-1}$ cm$^{-2}$, finding 
696 objects in six bands from $z = 0.08$ to $z =  0.4$.
\citet{hip03} 
surveyed 400 arcmin$^{2}$ to a flux limit of 3 $\times$ 10$^{-17}$ erg s$^{-1}$ cm$^{-2}$, 
finding 438 objects at $0.25<z<1.4$. Neither survey has extensive spectroscopic follow-up to give
unambiguous identification.
\cite{jon01} argued based on model luminosity functions that at their bright flux limits H$\alpha$ emitters would
dominate the counts; there was a small amount of overlap (18 of the brighter objects) with the Autofib redshift 
survey \citep{Ellis96}  which backed up their  conclusion. \cite{hip03} had
additional observations in a set of 15 medium and broad band filters which allowed the
redshifts to be constrained using fits to the spectral energy distributions. Spectroscopic followup of 
$\simeq 50$ objects with $R<25$ showed this was reliable for $\simeq$ 80\% of galaxies; other emission
line objects without continuum detections amount to $\lesssim$ 7\% of this sample (Hippelein 2004,
private communication). 

In this work we report on a small area (20 arcmin$^2$)
but extremely deep (1.7 -- 2.4 $\times$ 10$^{-17}$ erg s$^{-1}$ cm$^{-2}$)   survey carried out with the 
Taurus Tunable Filter on the
William Herschel Telescope in the Hubble Deep Field North (HDF-N) \citep{HDFN}. 
This field has the advantage of extensive
spectroscopy \citep{c96,c00}; so while the resulting sample is small it is unique in being 
almost completely spectroscopically identified. The goal
was to explore the possibility that additional star formation, which could contribute strongly to the 
cosmological SFR density, could be  occurring in galaxies with strong emission lines but little or no continuum emission.  

The plan of this paper is as follows: in Section 2 we review the details of the observation, data reduction, object identification and catalog making.  In Section 3 we discuss our process for 
calculating galaxies' SFRs and correcting for extinction. 
In Section 4 we describe the methodology we followed for determining the luminosity density based on the narrow band line fluxes we observed, and calculate the corresponding star formation rates. In Section 5 we present a
cosmic  star formation history, calculated from our sample 
to z=1, and discuss the implications.  Throughout this paper we use a cosmology:  H$_{0}$ = 70 km sec$^{-1}$ Mpc$^{-1}$, $\Omega_{m}$=0.3, and $\Omega_{\lambda}$ =0.7 motivated by \cite{Sper03}.

\section{Observations and Data Reduction}

The observations were carried out on 1997 March 11-14 at the 4.2m William Herschel Telescope (WHT) in the Canary Islands using the Taurus Tunable Filter \citep{TTF} on loan from the Anglo-Australian Observatory. The TAURUS2 instrument  \citep{TAURUS2} on the WHT was identical to the one on the Anglo-Australian Telescope.\footnote{Both TAURUS2 
instruments have now been decommissioned.} Operating at $f/4$ the instrument delivered a 
pixel scale of 0.29 arcsecs with the 1024$\times$1024 Tektronix CCD. Conditions were photometric for most of the run with
seeing of $\simeq 0.7 $arcsec.

The field centre was RA 12$^{\rm h}$ 36$^{\rm m}$ 46.8$^{\rm s}$ DEC +62$^\circ$ 14$'$ 46$''$ (J2000) and the WHT images including the whole of the Hubble Deep Field North together with its surrounding environs. By choosing a TTF etalon gap and 
broad-band blocking filter one generates a slice in wavelength  (which we call a `channel') to observe. We scanned the etalon in 3 contiguous wavelength regions, each a subset of the wavelength range defined by 3 blocking filters: I1 (7070\AA\ center/260\AA\ wide), I5 (8140/330\AA) and I8 (9090/400\AA). These are chosen to lie in the darkest regions of the night sky spectrum between 7000\AA\ and 10000\AA. The wavelength step for the scan was chosen to match, approximately, the Full-Width Half-Max of the etalon resolution. For etalon Lorentzian profiles this provides adequate sampling.  We scanned (i.e. changing the etalon gap to produce different filter wavelengths)  through 10 channels in the I1 and I8 filters and through 5 channels in the I5 filter. The full wavelength scanning parameters together with the exposure times per position are given in Table~\ref{tab:filters}. Longer exposure times were used in the redder filters, which correspond to higher redshifts
for a given line, to partially counteract cosmological dimming, The deepest scans were in the I8 filter where we exposed for 1.5 hours in each channel. 

\subsection{Data Reduction}

The reduction and analysis of tunable filter images is discussed in 
detail in \cite{JSB02}.
The goal in our case was to produce flux calibrated, 
background-subtracted data {\it cubes} (i.e. RA, DEC, $\Delta\lambda$). The raw CCD frames we de-biased 
and flat-fielded using dome flats. Frames taken at
the same wavelength were stacked with a cosmic-ray rejection filter. 
The overall wavelength calibration of the etalon gap was determined by taking a fast scan of an 
Argon arc lamp using just the central part of the CCD windowed. In the full frame data images the
Jacquinot spot is approximately centered on the CCD; 
the etalon phase shifts as one goes off-axis causing the tuned wavelength for a given channel to vary bluewards up
to an additional 10\AA. This
effect  also gives rise to a broad ring-shaped 
structures in the CCD background; the rings
correspond to the position of night-sky lines in the blocking filter band-pass. This was removed  by fitting for the centre of the ring pattern and then subtracting the median count in annular bins. The remaining background 
after this step still has structure due to CCD fringing; this is more random but had structure on $\sim 20$--30 pixel scales. We attempted to remove this by subtracting a local median, in a matching box, from each pixel. This procedure worked, essentially perfectly, in the I1 and I5 filters but was imperfect in the I8 filter due to the much greater fringing at 9100\AA. This gives rise extra non-Poisson noise which must be considered carefully when 
doing object detection (see Section~\ref{sec:det}).

The overall flux calibration was established by taking a fast windowed scan of a spectroscopic standard star and a bright reference star in the HDF-N field during photometric conditions. This was then referenced to the counts of the same reference star in the final data cube (which included some slightly non-photometric data)
in order to establish the final calibration of counts into flux versus wavelength.

\begin{figure*}[ht]
\epsscale{2.1}
\plotone{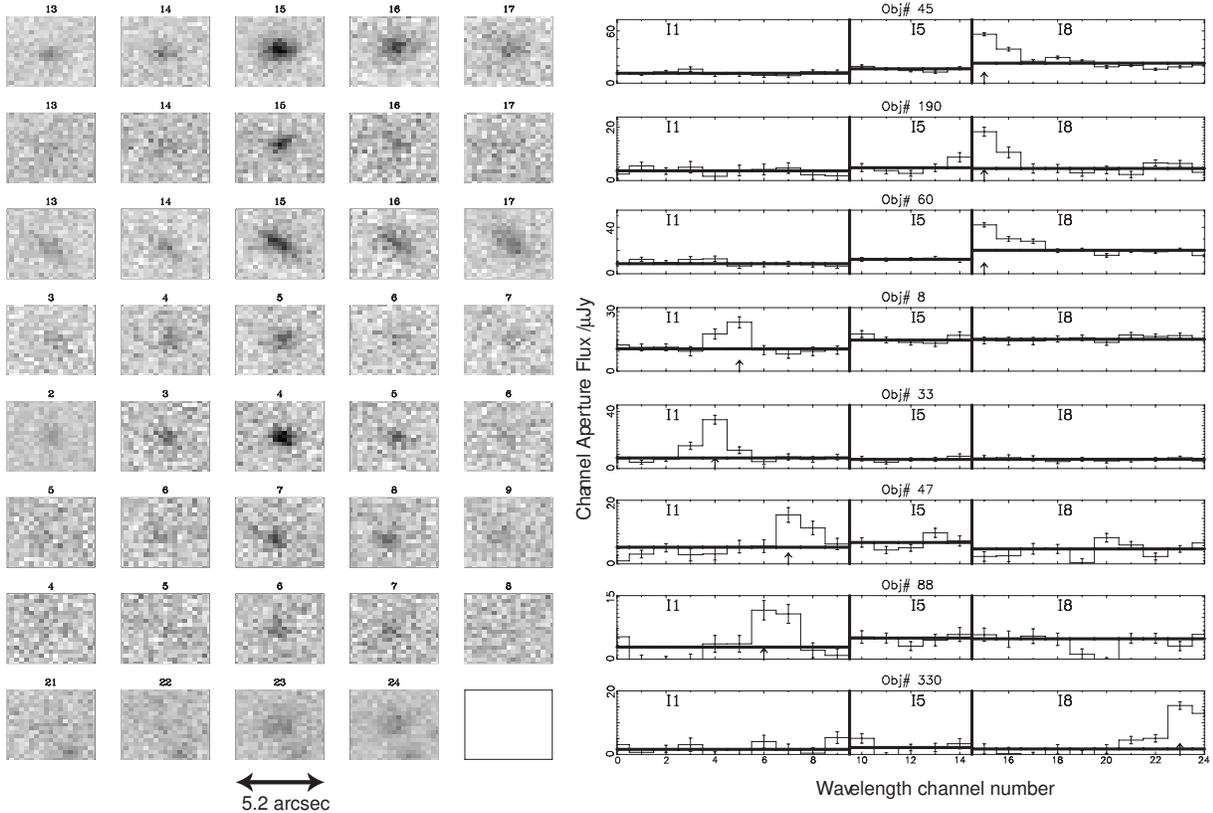}
\caption{Narrow band images and resulting spectra of our sample of 8 emission line
galaxies (top -- bottom, order matching Table~\ref{tab:objects}). The left panel shows the individual channel images (labeled with the channel number)  in the vicinity of the detected line, with the center column corresponding to the peak flux where the line is detected. The peaking of the flux is obvious, even in the 2D images. The right panel shows the full channel spectra, i.e. flux vs channel number (0--24). The arrow denotes the emission line location in each spectrum. The thick solid vertical lines represents the boundary between different filters where the wavelength scale is discontiguous. The thick solid horizontal lines denote the measured continuum level in each filter. Filter labels are given.  }
\label{fig:data}
\end{figure*}  

\subsection{Object detection and completeness tests} \label{sec:det}

Our goal was to construct a true three-dimensional catalog so we did not wish to require that an object have detectable flux in {\it every} wavelength channel. Rather if an object was detected in {\it one channel} we wanted
to have reliable flux limits in all other channels based on the aperture defined by the detection channel.
To achieve this we ran the SEXTRACTOR object detection
software \citep{sex} on each 2D channel {\it separately} using a threshold of at least 5 pixels $>$ 1.2 $\sigma$ above the background using pre-smoothing with a Gaussian kernel matched to the seeing. The noise $\sigma$
was set by the CCD readnoise and gain parameters. Certain parts of the 2D images ($\simeq 17\%$)
had residual contamination features due to optical ghosts and bad columns, these were masked off by a binary 
SEXTRACTOR mask to avoid object detection in these areas. The effective area searched for objects was then
20.35 arcmin$^2$. 

From each individual
channel catalog objects were photometered in all other channels using the same SEXTRACTOR aperture parameters (i.e. 'two-image' mode for all other channels as the second image). This process was then repeated
for each channel. What we end up with is a large set of catalogs, one for each channel, containing the {\it spectra} (i.e.
aperture flux versus channel) of each object based on it's detection in each channel. 
The next step was to merge these catalogs to remove duplicates where bright objects were detected in many channels. To detect duplicates a centroid proximity test of
$< 2$ pixels was used. To remove duplicates we kept only the spectrum based on the channel with the strongest detection. Some additional cosmic-ray filtering was performed at this step; essentially removing
any source too sharp to be a real object. The result of the merging procedure
was a  master catalog of all possible objects (351 total) in 
all possible channels with associated spectra and noise vectors. This 3D approach is very similar to that of
\cite{jon01} but differs from \cite{hip03} who used the sum of the channel images for object detection. For objects which
only appear as a line in one channel the latter approach is less sensitive.

The final step is to search for emission lines in the final spectra. Choosing the right signal:noise cut is critical to avoid contamination by spurious objects. One test we performed was to compare the actual variance in the spectra with that calculated based on CCD gain and readnoise. (The majority of objects will not have 
a spectral feature in the observed wavelength range as it is so short and so the variance is dominated by
noise). We found these agreed well in the I1 and I5 channels but fell short by a factor of 2 in the I8 channel. This 
was because I8 had additional channel-to-channel fluctuations due to significant residual fringing. To compensate for this we forced a match by doubling the calculated noise in the I8 channel. We estimate the final 1-$\sigma$ noise in each channel to be $(7.9,6.0,5.5)\times 10^{-18}$  ergs cm$^{-2}$ s$^{-1}$ for I1, I5 and I8 respectively. 

For emission line detection we smoothed the spectra with a 3 pixel kernel (values $[0.33,1,0.33])$ 
in order to approximately match the Lorentz profile of unresolved lines). The continuum, if present, was defined by the median flux value in each filter. A line was `detected' if any pixel had flux $> 4.5 \sigma$ above the continuum, this threshold being
equivalent to 3$\sigma$ in the unsmoothed data. This gave a `3$\sigma$' emission line
catalog of 41 `objects'. Inspection of the images showed that many of these objects looked more like artifacts than real objects; the effect of the fringing
being to impart a non-Gaussian characteristic to the noise.We decided to calibrate the `believability threshold`
empirically by searching for emission line in 1000 randomized sky positions (constrained to be distant from 
all known objects in the field).  This let us estimate how often by pure chance, given the skew noise distribution
in our images, we would incorrectly detect an emission line when photometering our data. Spectra were extracted in 3 arcsec diameter apertures and the same 
detection algorithm applied to measure the false-positive rate. 
At 3$\sigma$ the contamination due to non-Gaussian artifacts is severe: in our master catalog we calculated
a rate of 40 false positives for every 351 random positions in our synthetic catalog. Since the actual catalog of 
351 objects had 41 emission line detections this led us to conclude that at this $S/N$ threshold 
we are overwhelmingly dominated by false objects. Using these simulations as a guide we raised the 
threshold to make a `5$\sigma$' catalog of 8 objects --- the random catalog showed in this we expect
on average only 0.7 objects to be false. This rapid drop-off in the number of `objects' in the random catalog 
when moving from 3$\sigma$ to 5$\sigma$ (only a factor of 1.66 in flux) is consistent with the objects in the
former catalog being noise artefacts and not serendipitous galaxies. Visual inspection
confirmed that these 8 objects were very likely genuine galaxy images. We show images and spectra of all 8 objects in Figure~\ref{fig:data} and tabulate them in Table~\ref{tab:objects}.

\subsection{Redshift identification}

Redshifts were confirmed by cross-matching the sky-coordinates with published HDF spectroscopy from \cite{c96,c00}.
 In 7 out of the 8 cases there was a clear, positive, unique identification where a known emission line galaxy at the correct RA,DEC had the correct redshift to put a strong line in the right filter. (The probability of the last happening by chance is of order 1\%). Out of these 7 objects one is in the HDF itself (\#8) and the other 6 are in the flanking fields.
The 8th object is also in the flanking fields and is blended with a neighboring object
in ground-based images of \cite{cap04}. Its identification remains a mystery but we note the object is
detected in the $B$-band so it is unlikely to be $z=5.5$ Ly$\alpha$. We further note the peak flux is at the red end of the I8 scan so it could correspond to a continuum break instead of an emission line.

The final redshifts and line identifications are given in Table~\ref{tab:objects}. Figure~\ref{fig:data} shows both the data cube
images of each object in the vicinity of the line and resulting channel spectra.  The emission lines are clearly evident in both.
Continuum was also significantly detected 
in these objects and the flux levels agree with the published $I$-band HDF-N photometry of the same objects. 
The magnitude range of the objects is $18.0<I<20.2$. Since the number of objects was small and the magnitudes were relatively bright our main result --- no surprises --- is presaged here. However the line flux limits reached are quite faint (see above and 
Table~\ref{tab:sfr}): in these redshift windows probed this sample represents all the objects down to a small
fraction of the Schechter luminosity $L^*$.

\begin{table*}[ht!]
\caption{Etalon scan parameters in the three different filters. Each wavelength range  and emission line identification defines a redshift range and corresponding volume, shown below. Volumes are in units of Mpc$^{3}$.}\label{tab:filters}
\footnotesize 
\begin{center} 
\begin{tabular}
{     c     |      c  |  c       c      |      c       c     |        c         c     }
 \hline
 \hline     
            &                &      &                 &                &                   &              &         \\[-3mm]
 Parameters of    & Exp/channel  & \multicolumn{2}{c}{H$\alpha$}  &    \multicolumn{2}{c}{H$\beta$}    & \multicolumn{2}{c}{[OII]}\\
 Etalon scan&   (Secs)  &  z           &  Volume         &      z         &   Volume          &         z    & Volume  \\
 \hline     &        &            &                 &                &                   &              &         \\[-1mm]
            &      &              &                 &                &                   &              &         \\[-5mm]
 7058--7112\AA, 10 steps   of & 800  & 0.075--0.084      & 7              &0.452--0.463       & 197               &0.894--0.908     & 599     \\
$\Delta\lambda=$ 6.0\AA,  FWHM =  6.3\AA          &      &              &                 &                &                   &              &         \\ [2mm]
 8100--8133\AA, 5 steps   & 1800  & 0.234--0.239      & 31              &0.666--0.673       & 203               &1.173--1.182   & 462    \\
 of $\Delta\lambda=$ 8.3\AA, FWHM =  8.9\AA     &         &                &                 &                &                   &              &         \\[2mm]
 9041--9125\AA, 10 steps of& 5400   & 0.378--0.390      & 176             &0.860--0.877       & 696               &1.426--1.449   & 1340    \\
   $\Delta\lambda=$ 9.3\AA,  FWHM =  13.5\AA    &           &                &                 &                &                   &              &         \\
 \hline
\end{tabular}
\end{center}
\end{table*}

\section{Line luminosity determination and Extinction Corrections} \label{sec:lines}

Our method to calculate SFRs is to calculate from each observed line luminosity the equivalent, extinction-corrected, H$\alpha$ 
luminosity and then convert H$\alpha$ to SFR using a constant converstion ratio which we assume is independent of
redshift (this is very similar to assuming a constant IMF).

To standardize our extinction correction across the three different observed lines, we converted the H$\beta$ and [OII] fluxes to equivalent H$\alpha$ fluxes, then corrected the H$\alpha$ fluxes for extinction.  We have no
alternative but to  assume a mean extinction in order to make progress. To dust-correct
the H$\beta$ flux, we use the relation derived by \citet{tre98} which covers a similar redshift range to our H$\beta$ and
H$\alpha$ galaxies:
\begin{equation}
\frac {F_{H\alpha,raw}}{F_{H\beta,raw}} = 2.86 \times 10^{-C(-0.323)} 
\end{equation}
where we use their mean $C=0.46$, which correspondes to $A_v=CR/1.47 = 1.0$ mag assuming R=3.2 \citep{sea79}.  

For [OII], we use the {\it observed\/} ratio:
\begin{equation} 
\frac{F_{H\alpha,raw}}{ F_{OII,raw}} = 2.22
\end{equation}
 from \citet{ken92}. There may be more complex dependencies of this ratio on, for example, luminosity or metallicity (e.g. \cite{jan01}). 
 In particular more recent analyses of much larger local samples \citep{jan01,hop03} have found  $F_{H\alpha,raw}/F_{OII,raw}$ = 4.34).  Since we  have one [OII] object in our sample it is only necessary to have one conversion factor, we choose to use the \cite{ken92} value. This also agrees with the $z=1$ value found by \cite{kgb99}; however that work concerned luminous $L^*$ galaxies
 whereas our object is somewhat sub-$L^*$. If a higher ratio was adopted this would increase our highest redshift measurement of the star-formation rate density (below) by a factor of two. Our values are currently at the low-end of observed values and so this would bring the results more in to agreement. However given the large error bar from one object this point is somewhat moot.

To correct the equivalent H$\alpha$ fluxes for extinction, we adopt the $A_V=1.0$ mag value giving 
\begin{equation}
  F_{H\alpha,corrected}= 2.12 \times  F_{H\alpha,raw} \label{foo1}
\end{equation}

The raw and corrected equivalent H$\alpha$ luminosities are shown in Table 2. We then use the 
 \citet{ken98} conversion SFR $=$   L(H$\alpha$) $\times$ $7.9\times 10^{-42}$ $M_{\sun}$ $\rm yr^{-1}$ 
$/$ ergs cm$^{2}$ s$^{-1}$ (which assumes a \citet{sal55} IMF) to convert to a star formation rate.  

\begin{figure}[t]
\epsscale{1}
\plotone{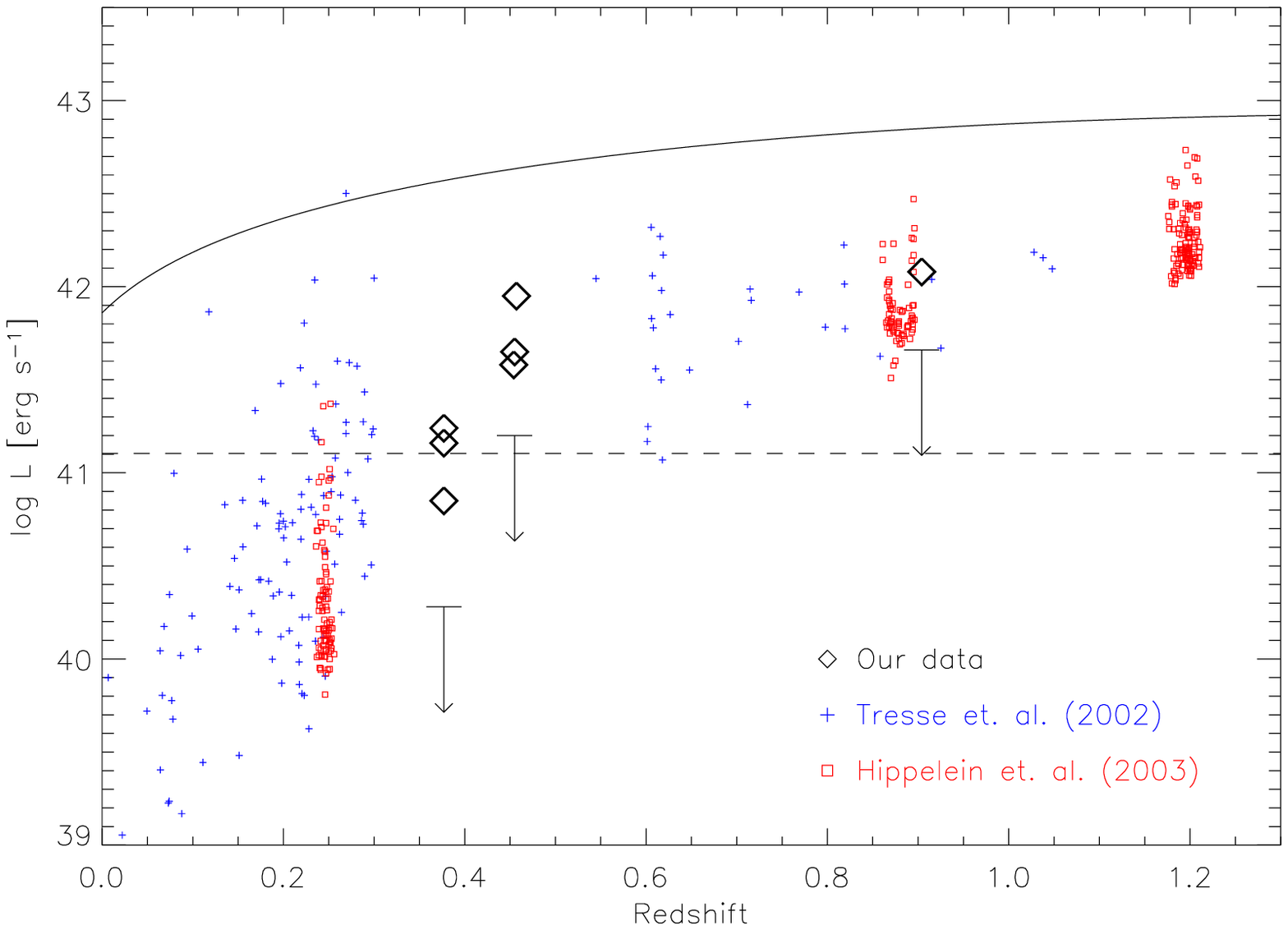}
\caption{Equivalent H$\alpha$ luminosity (dust-corrected) versus redshift of individual galaxies. Also shown are lower limits calculated from our flux limits. The solid line shows for reference the {\em shape} of the cosmological SFR density-$z$ curve as parameterized by \cite{Col01} (extinction corrected) re-normalized to an H$\alpha$ luminosity of $10^{41.86}$ ergs s$^{-1}$ at $z=0$ (this is the
value of $L^*$ quoted by \cite{gal95}).  The dashed, horizontal line shows the equivalent luminosity corresponding to SFR $=$ const. $=1$ $M_{\odot}$yr.$^{-1}$  (We assume no redshift dependence of the Kennicutt $L(H\alpha)/SFR$ conversion.)}
\label{fig:sfr-ind}
\end{figure}

 Figure~\ref{fig:sfr-ind} shows how 
the resulting line luminosities we measure for individual galaxies compare with others in the literature ([OII] and
H$\alpha$ luminosities from \cite{tre02}, \cite{hip03}, which we
also standardize to dust-corrected H$\alpha$ luminosities in the same way as for our data). 
Our objects are all only moderately star-forming, typically only a few $M_{\odot}$ yr$^{-1}$  and our detection limits correspond to quite low luminosity values.

\begin{table*}[ht!]
\caption{Raw and extinction-corrected H$\alpha$ equivalent luminosities of each galaxy.  Luminosities are shown as log L in units of erg s$^{-1}$. }\label{tab:objects}
\begin{center}
\begin{tabular}
{    c   |  c     |       c      |      c       |          c      |         c   |       c    }
 \hline
 \hline     
   &         &                &                 &            &    &     \\[-3mm]
   &         &                &      Raw Line        & Corrected  &    &     \\
 Obj\#   & z         &    Line        &   Luminosity    &H$\alpha$ Luminosity   & RA & Dec    \\
 \hline  &   &                &                 &            &          &             \\[-1mm]
        &    &                &                 &            &          &             \\[-5mm]
 45   & 0.378       &  H$\alpha$     &       40.94     & 41.27      &12 37 04.65&62 16 52.1\\
 190 & 0.378       &  H$\alpha$     &       40.53     & 40.85      &12 37 04.04&62 15 23.2\\
 60  &  0.378       &  H$\alpha$     &       40.83     & 41.16      &12 36 39.72&62 15 26.2\\
 8    & 0.454       &  H$\beta $     &       40.65     & 41.58      &12 36 42.93&62 12 16.3\\
 33  & 0.457       &  H$\beta $     &       41.02     & 41.95      &12 36 58.39&62 15 48.7\\
 47  & 0.455       &  H$\beta $     &       40.71     & 41.64      &12 36 31.16&62 12 36.2\\
 88  & 0.904       &  [OII]           &       41.41     & 42.08      &12 36 38.80&62 15 47.2\\
 330 & ?             &  ?                &    ?                &    ?            & 12  36  43.18 & 62  16  25.1 \\  
 \hline
\end{tabular}
\end{center}
\end{table*}
  
\section{Calculation of luminosity and SFR densities}

To calculate SFR densities (i.e. SFR per unit comoving cosmological volume) we use the same SFR/luminosity
conversion procedures but this time applied to the dust-corrected H$\alpha$ luminosity density.
Because our flux limits are well below $L^*$ at all redshifts the very simplest estimator for the luminosity density $\rho_L$:
\begin{equation}
 \rho_L(z_k) = \sum_i \frac{L_i}{V_k}   \label{eqn:direct}
\end{equation}  
where $V_k$ is the volume in slice $k$ given in Table 1 comes very close
to estimating the total luminosity density
irrespective of the shape of the luminosity function at the faint end. 
(Note we are using dust-corrected equivalent H$\alpha$ luminosities). Any new population
of faint line emitting galaxies would show up as an excess luminosity density
compared to other surveys.

In order to estimate errors and the amount of missing light below our flux limits we next 
assume the $H\alpha$ luminosity function takes the form of a \cite{Sch76}  function  
\begin{equation}
\phi(L)dL = \phi^* (L/L^*)^\alpha exp(-L/L^*)d(-L/L^*) \label{eqn:sch}
\end{equation}  
and estimate $\phi^*, L^*$ as follows.  Writing $x=L_{lim}(z_k)/L^*$ 
where $L_{lim}(z_k)$ is the luminosity limit corresponding to the flux limit in a 
given slice,  the
number of galaxies per unit volume $n$ we expect to see is:
\begin{equation}
n =  \phi^*  \int_x^\infty x^\alpha e^{-x} dx = \phi^*    \Gamma(\alpha + 1,x) \label{eqn:n}
\end{equation}  
where $\Gamma$ is the incomplete Gamma function. Similarly 
the integrated luminosity density $\rho_L$ above the flux limit is

\begin{equation}
\rho_L =  \phi^* L^*  \int_x^\infty x^{\alpha+1} e^{-x} dx = \phi^* L^*   \Gamma(\alpha + 2,x)  \label{eqn:rho}
\end{equation}  
Then we can estimate $L^*$ using:
\begin{equation}
\frac {\Gamma(\alpha+1,x)}{L^*\Gamma(\alpha+2,x)} = \frac {n}{ \rho_L} \label{eqn:rat}
\end{equation}
which can be solved iteratively for $L^*$ given the observed values of $n$ and $\rho_L$ in
a given redshift slice (in effect we constrain $\phi^*$ and $L^*$ to give the observed luminosity
density $>$ our flux limit, but they do play a role in the error calculation below). 
Then $\phi^*$ can be derived using equation~\ref{eqn:n} and 
the {\em total\/} luminosity density calculated by putting $x=0$ in equation~\ref{eqn:rho}.

The amount of light missed below our flux limits depends mainly on $\alpha$. For our final, `total`, values we adopt
$\alpha$ of $-1.3$ which is the mean of local optical--IR surveys \citep{gal95,blan01,huang03}. With this value we find
that we are seeing between 89\% of the light in the $z=0.38$ bin to 61\% of the light in the highest $z=0.90$ bin. If we 
$\alpha$ were higher we would of course miss more. For example if we used $\alpha=-1.6$, which is seen in the ultraviolet
luminosity function of $z>3$ star-forming Lyman break galaxies \citep{stei99}, then our corresponding luminosity
completeness is only 76--43\%. 

\begin{table*}[ht!]
\caption{Values calculated in determining the star formation rate density. Luminosities refer to
equivalent dust-corrected H$\alpha$ values derived from each actual line. }\label{tab:sfr}
\footnotesize
\begin{center} 
\begin{tabular}
{      c | c | c     |       c      |      c       |          c        |      c        |      c   }
 \hline
 \hline     
  &   &        &                &                 &                &                       &  \\[-3mm]
  &     &  Log     &       Log      &      Log        &     Log        &       Log             &  Luminosity\\
 $z$ &  Line    &  $\sum_i L_i / V $    &    $\phi^{*}$  &        L$^{*}$  &     $\rho_{L}$ (TOTAL) &    $\rho_{SFR}$       &  Function\\
     & & (erg sec$^{-1}$ Mpc$^{-3}$)   &  (Mpc$^{-3}$)  & (erg sec$^{-1}$)&(erg sec$^{-1}$ Mpc$^{-3}$)& (M$\sun$ yr$^{-1}$ Mpc$^{-3}$)& Limit   \\
 \hline  &   &  &                &                 &                &                       &   \\[-1mm]
      &      &  &                &                 &                &                       &   \\[-5mm]
 0.38 & H$\alpha$  & 39.36 &  -2.55       &       41.82     &   39.41        &   $-1.70^{+0.14}_{-0.21}$              &  .03 L$^{*}$\\
       & &  &                &                 &                &                       &   \\[-2mm]
 0.46 & H$\beta $  &  39.94  &  -2.23       &       42.15     &   40.07       &    $-1.04^{+0.17}_{-0.14}$              &  .11 L$^{*}$\\
     &    &  &                &                 &                &                       &   \\[-2mm]
0.90 & \hbox{[OII]} &   39.30   &    -2.91       &       42.30     &   39.52        &    $-1.58^{+0.34}_{-0.30}$              &  .23 L$^{*}$\\
     &   &  &                &                 &                &                       &   \\
 \hline
\end{tabular}
\end{center}
\end{table*}

Errors are tricky to calculate 
in samples with small numbers of objects \citep{geh86}, we adopt a Monte-Carlo approach
where we  populate a model universe according to the Schechter function (equation~\ref{eqn:sch}). This lets
us account for scatter both in the luminosities and the numbers of objects in a finite survey. A grid
of $(\phi^*, L^*)$ values are explored bracketing our measured values, 
for each grid point we make 1000 realizations of our observed
volume. For each realization we repeat our estimation of the luminosity and number densities using 
Equations~\ref{eqn:n}--\ref{eqn:rat}. We can then compute a mean 
$(\bar\rho, \bar n)$ and variance $(\sigma_\rho^2, \sigma_n^2)$  for these quantities and 
form a $\chi^2$ value to characterize the goodness of fit of the grid point to the observed data $(\rho,  n)$:

\begin{equation}
\chi^2 = \left( \frac{\rho - \bar\rho}{\sigma_\rho} \right)^2 +   \left( \frac{n - \bar n}{\sigma_n} \right)^2
 \label{eqn:chi2}
\end{equation}  

Then the range of luminosity densities corresponding to the $\Delta\chi^2=1$ contour gives us our error bars.
Note these are not symmetric and the direction of the asymmetry changes with redshift. This is because our
$\chi^2$ statistic has both luminosity and number components whose relative importance changes; this
is compounded with the high asymmetry of the Schechter function. We note that for the $z=0.90$ bin where
there is only one object the uncertainty is a factor of two which is in accord with intuitive expectations.
The final luminosity densities along with our calculated $\phi^*$ and $L^*$ are listed in Table~\ref{tab:sfr}. We also tabulate  the results of the simple direct estimate (equation~\ref{eqn:direct}) as well as our more sophisticated method.

\begin{figure}[ht]
\epsscale{1}
\plotone{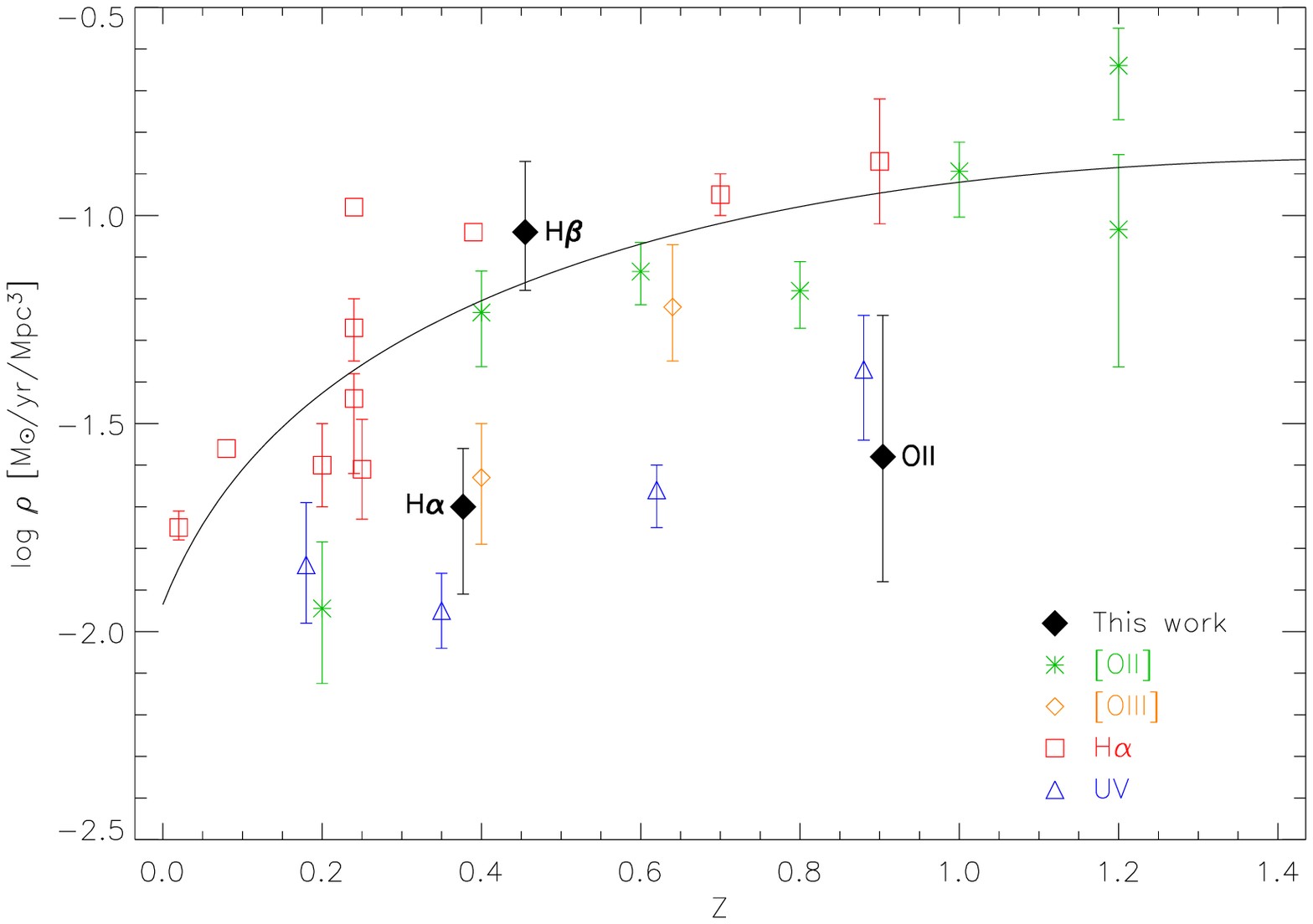}
\caption{Star formation rate density versus redshift.  [OII] data is from \citet{hog98} and \citet{hip03}; H$\alpha$ data is from \citet{gal95}, \citet{tre98}, \citet{kgb99}, \cite{jon01}, \cite{pas01}, \citet{tre02}, and \citet{hip03};  [OIII] data is from \citet{hip03}; UV data is from \citet{lil96} and \citet{try98}.  The solid line shows for reference the fit to the cosmological SFR density-$z$ data of \cite{Col01} (extinction corrected version).}
\label{fig:sfrd}
\end{figure}

The same \citet{ken98} conversion factor from in Section~\ref{sec:lines}
converts luminosity densities in to SFR densities. 
The resulting SFR densities are given in Table~\ref{tab:sfr} and in 
Figure~~\ref{fig:sfrd} are compared with others in the literature from both broad-band UV estimates and narrow line estimates (for the latter we go back to the original data and consistently
use our conversions/extinctions for line luminosity densities to SFR densities). In general our results fall within the scatter of previous measurements given
the large error bars from the small number statistics of our sample. Our H$\alpha$, H$\beta$ points agree with the range of Balmer line estimates at $z<0.5$ from the literature, at $z=0.9 $ our [OII] estimate is a little low 
(even given the large one object error bars) especially compared to 
other [OII] estimates. It is possible that our measure is underestimated due to an incorrect choice 
of [OII]$/$H$\alpha$ ratio, for example if we adopt the `unbiassed' [OII]$/$SFR conversion of \cite{RG02} we can
raise our [OII] point by 0.35 dex. It is also possible that our assumed extinction value is incorrect. The effect of increasing the extinction, for example to 
$A_V = 2.0$ mags would be greatest on our $H\beta$ SFR density raising the estimate by 0.5 dex while the other estimates would be raised by 0.3 dex. This would make [OII] agree better but then the H$\beta$ point would be highly discrepant. Similarly the effect of assuming zero extinction would be to lower the points by the same amounts. Clearly then our SFR density measurements would still be inconsistent: [OII] would be highly discrepant. Internal consistency gives us good reason to think our choices of [OII]$/$H$\alpha$ line ratio and extinction are reasonable, though the allowed range is large. 

The formal error bars also do not include the effect of large scale structure which are significant. Our small volume is
equivalent to a sphere of radis 5--6 Mpc which is 
$\simeq$ the local blue galaxy correlation length. (\cite{Mad03} finds  $r_0=5.2$--5.9 Mpc for star-forming galaxies.)
For the volumes ($\sim$ 500--1000 Mpc$^3$)  and number densities (0.002--0.02 Mpc$^{-3}$) considered here we expect fluctuations of 100--130\%
relative to the mean using the results of \cite{Somer04}. 
The deviations of the points from the line in Figure~~\ref{fig:sfrd} are of this order 
and so large scale structure would represent an equally reasonable explanation of our scatter.

\section{Conclusions}

This work demonstrates we can select galaxies purely by line emission in a three-dimensional survey, producing a pure narrow-band sample limited by line luminosity to constrain the cosmological SFR.  
One can see from Table 3 that our observable luminosity limit for $H\alpha$ is almost two orders of magnitude lower than  $L^*$, indicating a small completeness correction.  The luminosity densities calculated from this data are therefore very complete and provide a strong bound on the cosmological total.
  
Our study also finds no evidence for any new population of low-luminosity objects with strong emission lines which could have caused previous estimates of the SFR density to be seriously underestimated. Searching for such unexpected
populations was a primary goal of our observations, we are reporting a null result for this search.

Rather we find a SFR density history consistent (albeit with large errors from small number/volume
statistics) with earlier estimates from broad-band selected samples. More
specifically our measured SFR density agrees with the previous emission line determinations (both broad-band and
narrow-band selected) but are higher than previous ultraviolet determinations which is in accord with earlier findings.
Our low sub-$L^*$ line luminosity limits corresponds to 
seeing $\simeq$ 80--90\% of the integrated SFR density at $z\simeq 0.4$; even at $z=1$ our flux limit corresponds to $\simeq$ 40--60\% of the total star-forming light. This indicates that previous completeness estimates to broad-band surveys were correct and that we now have a reasonable census of the total star-formation rate density in the Universe out to $z=1$.

The success of narrow-band tunable filter surveys (both this work and those of \cite{jon01} and \cite{hip03}) augurs well for the next generation 
of tunable filter 3D instrumentation. Galaxies can be located in data cubes purely from their line emission {\em flux}
(i.e. not equivalent width) and recovered 
aperture-extracted spectra of general objects agree well with those taken using classical dispersive spectrographs. For line emitting objects modest exposures on 4m telescopes allow one to survey substantially below $L^*$ in volume-limited samples at high redshift. Our survey suffers from a limited volume surveyed using an instrument with a 
5 arcmin FOV; future wider field instrumentation should allow much bigger surveys. 
In the  near-future we also expect such instrumentation techniques to be extended beyond the CCD cutoff
to the near-IR and especially the $J$-band.
Here the gain by observing in narrow wavelength slices is even greater than in the optical because the airglow lines are even stronger. Here tunable filter techniques \citep{Joss04} on 8m class telescopes
would allow astronomers to find Ly-$\alpha$ emitters at $z>7$ for even very modest
Ly-$\alpha$ $/$ SFR assumptions and thus allow us to experimentally probe the very  first galaxies and
the epoch of re-ionization.

\bigskip

\acknowledgments {\bf Acknowledgements}
Based on data from the William Herschel Telescope which was operated by the Royal Greenwich Observatory in the in the Spanish Observatorio del
Roque de Los Muchachos of the Instituto de Astrofisica de
Canarias. We would especially like to thank the telescope staff for their excellent support.
KG \& JT acknowledge generous support from the The Johns
Hopkins University and the David and Lucille Packard Foundation.  ST acknowledges
a summer studentship  from the Anglo-Australian Observatory. 
Travel support for KG \& JBH  was provided by 
the Australian Nuclear Science and Technology Organization. Roberto Abraham acknowledges support from
the U.K. Particle Physics and Astronomy Research Council and the Royal Greenwich Observatory, Cambridge.

\end{document}